\documentclass{PoS}

\usepackage{lineno}

\title{HAWC: Design, Operation, Reconstruction and Analysis}

\ShortTitle{}

\author{\speaker{Andrew J. Smith}$^a$ for the HAWC Collaboration$^b$ \\
        \llap{$^a$}Department of Physics, University of Maryland, College Park, MD 20742, USA.\\
        \llap{$^b$}For a complete author list, see \href{http://www.hawc-observatory.org/collaboration/icrc2015.php}{www.hawc-observatory.org/collaboration/icrc2015.php}. \\
        Email: \email{asmith8@umd.edu} }

\abstract{The High-Altitude Water Cherenkov (HAWC) Observatory was completed and began full operation on March 20, 2015. The detector consists of an array of 300 water tanks, each containing ~200 ktons of purified water and instrumented with 4 PMTs. Located at an elevation of 4100m a.s.l. near the Sierra Negra volcano in central Mexico, HAWC has a threshold for gamma-ray detection well below 1 TeV and a sensitivity to TeV-scale gamma-ray sources an order of magnitude better than previous air-shower arrays. The detector operates 24 hours/day and observes the overhead sky (~2 sr), making it an ideal survey instrument. We describe the configuration of HAWC with an emphasis on how the design was optimized, describe the data acquired, reconstructed and analyzed. Finally, we will demonstrate the sensitivity of the detector using the observation of the Crab. This paper serves as a detailed technical description of the foundations of the numerous analyses presented at this meeting by members of the HAWC collaboration.}

\FullConference{The 34th International Cosmic Ray Conference,\\
		30 July - 6 August, 2015\\
		The Hague, The Netherlands}

\begin{document}

\section{Introduction}

The field of ground-based gamma-ray astronomy has rapidly evolved in the last quarter century since the discovery of the Crab nebula
by the Whipple collaboration~\cite{whipple}. At TeV energies, non-thermal processes dominate emission, so the detection of these particles
acts as a probe the most fascinating an extreme processes in the universe, including behavior of compact objects, neutron stars and
black holes, the death of stars through supernova, gamma-ray bursts and the potential indirect detection of Dark Matter particles.
The field's rapid expansion was mainly driven by the utilization of Imaging Atmospheric
Cherenkov Telescopes (IACTs), which are pointed optical telescopes that image the Cherenkov light produced in the atmosphere by 
electromagnetic cascades (also called showers) initiated by high-energy particles. An alternative method for detecting high-energy 
gamma-rays is to detect the shower particles directly at the ground level using arrays of particle detectors called Extensive
Air Shower (EAS) arrays. This second type of instrument offers some distinct advantages over IACTs as they can operate
at continuously, not just on dark clear nights, and can observe the entire overhead sky.  This paper describes the High Altitude 
Water Cherenkov (HAWC) a second generation EAS observatory, which recently began operation.

The HAWC Observatory is a new and novel TeV gamma-ray detector located on a high mountain in central Mexico. 
It's sensitivity is about an order of magnitude better than the preceding projects, such as Milagro and ARGO, and for 
long exposures rivals that of pointed IACTs, especially above a few TeV. Though it has only about 1/20th the instantaneous 
sensitivity of these other TeV instruments, the wide field of view (2 steradians) and 24/7 operation gives HAWC greater than 1000 times 
the exposure compared to narrow field pointed instruments. This contrast in capabilities makes HAWC and IACT observations complementary, 
HAWC being ideal for daily monitoring, transient detection due to its high duty factor while IACTs are ideally suited for deep pointed observations.

In this paper, we will describe the design of HAWC, stressing the optimization of the design choices. Then we will briefly describe the data produced 
by HAWC and how it is reconstructed and processed. Finally, we will discuss the potential for improvement of the technique and the importance and 
further potential of gamma/hadron separation using the water Cherenkov technique.

\section{Design Considerations}

At the TeV scale, the brightest astrophysical gamma-ray sources (such as the Crab) produce only about 1 particle (E>1TeV) per million square 
meters per second. To obtain reasonable event rates, an instrument must have a collection area of > 10,000 m$^2$. Unlike 
with IACT's, which can image a large area of atmosphere with an array of relatively small telescopes, an the effective area 
of an extended air-shower array (EAS) is at most approximately equal to its physical size. HAWC covers an area of ~22,000 
m$^2$ with more than half of that area is covered particle detectors.

Achieving the lowest possible energy threshold is also critical to optimizing sensitivity,
since VHE sources have spectra which rapidly decline with energy. This is achieved 2 ways: 1) locate
the detector at an extreme elevation and 2) design an instrument which has a high density and a
high efficiency for shower particle detection. To address (1), we situated HAWC at a site with
an elevation of 4100m a.s.l. (~630 gm/cm$^2$). This altitude is about at the limit of where 
humans can function efficiently without
supplemental oxygen, which could drastically increase construction costs. At this elevation, 
~10\% of the energy from a 1 TeV gamma-ray shower will reach the detector level, which is about 
5x the energy which reaches the Milagro elevation (2600m) and more than 25x the energy that reaches 
sea level. To address (2) HAWC 
utilizes water as both a detector medium for charged particles, through the Cherenkov light they 
produce, and as a converter medium that stops the gamma-rays in the EAS shower, which 
carry ~80\% of the shower 
energy, to electrons where they can be detected. Each of HAWC's 300 water tanks has an area 
of 42 m$^2$.

Finally, to achieve the highest possible sensitivity, the enormous cosmic-ray backgrounds must
be efficiently suppressed. The rate of hadronic background events above 1TeV is  ~0.1 
particle/$m^2$/$s$/$sr$. Even restricting observations to a small 1 deg$^2$ bin, the of hadronic
background is about 1000x greater than that of the brightest gamma-ray sources.
HAWC's trigger rate purely from cosmic-rays is 10-20 kHz. 
These enormous background rates limit the gamma-ray sensitivity of any instrument unless 
they can be attenuated. IACTs have found that the shape of the shower images, which is a direct 
measurement of how energy is distributed in the shower,  
is a strong indicator of progenitor species. This shape imaging is also possible in HAWC 
so long as the water tanks are deep enough to act as effective calorimeters.
To achieve this, it is important that the tanks are sufficiently 
deep enough to absorb and reliably measure the energy of EAS particles reaching the ground level.
This requires that the water depth be much larger than the radiation length. We chose a depth
of 4.0m (water above PMTs), about 10 radiation lengths. This deep water volume also permits us to detect 
large signals from thru-going minimum ionizing muons, which are common in hadronic showers and 
have long tracks that produce large Cherenkov light signals.

\section{Technical Description}

\begin{figure}
\includegraphics[width=3in]{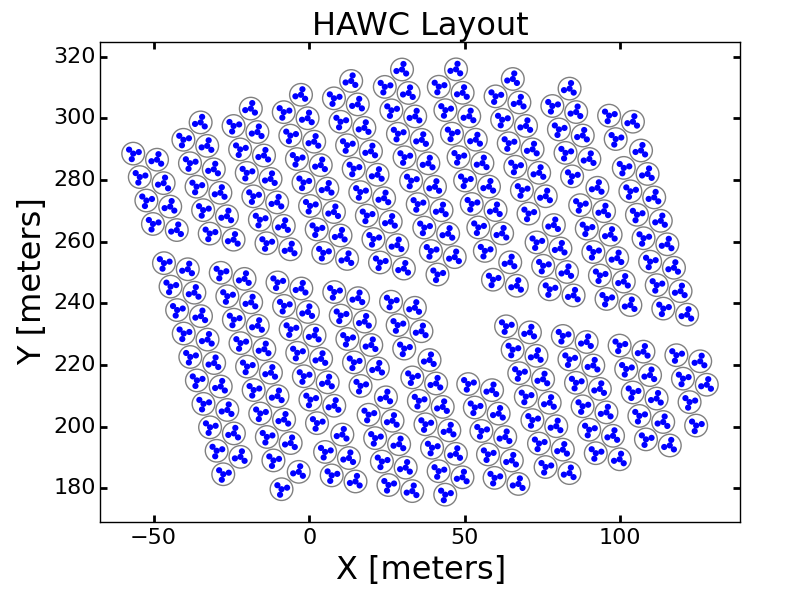}
\includegraphics[width=3in]{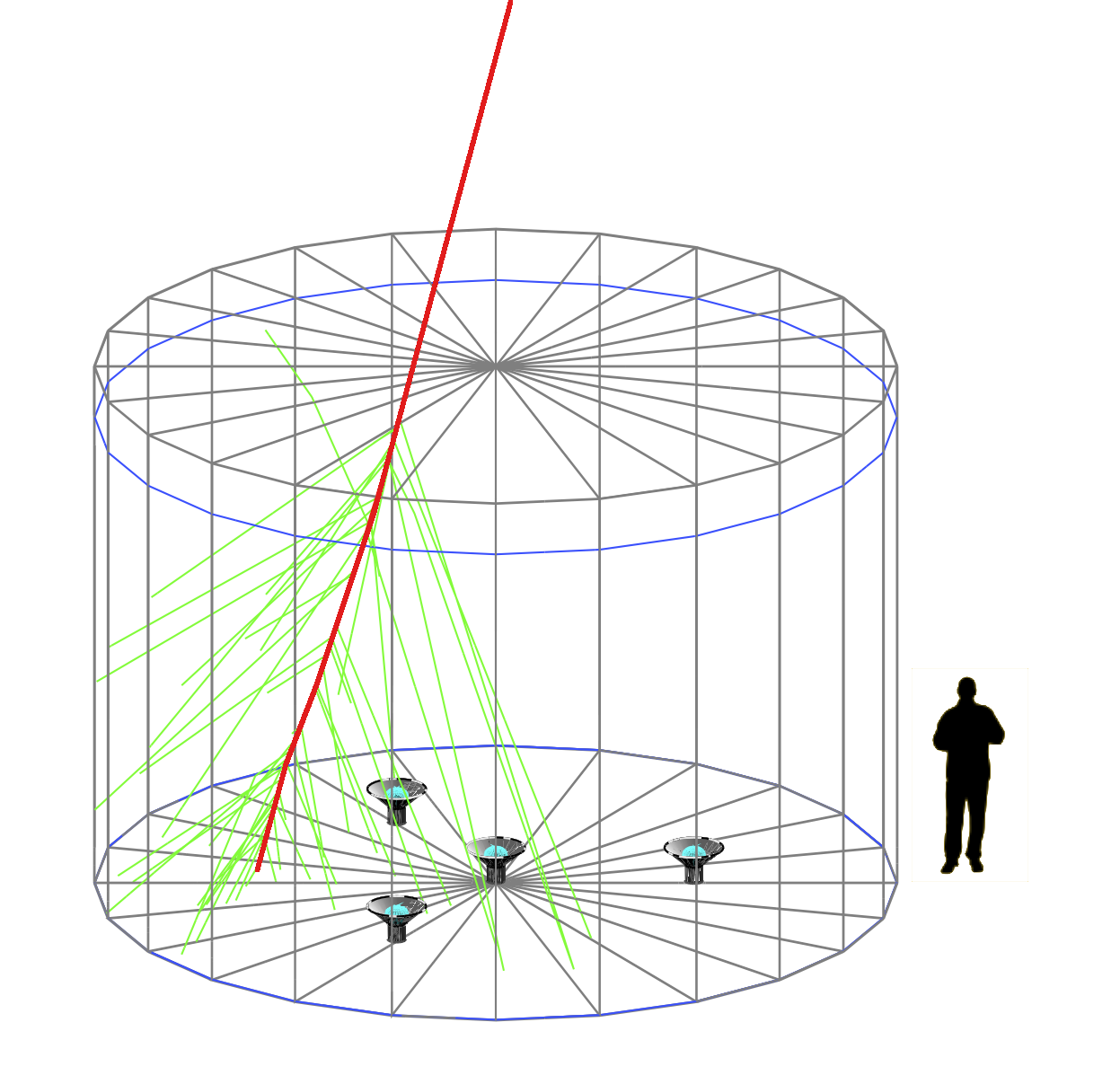}
\caption{Layout of the HAWC array (left) and a diagram of a single HAWC tank (right).}
\label{fig:layout}
\end{figure}

HAWC is constructed from 300 identical 7.3m diameter commercial water storage tanks. The layout
of the HAWC tanks is shown in figure \ref{fig:layout}.  The tanks
are assembled from 5 rings and each ring is made up from 8 corrugated steel panels. The panels
are light enough to carried workers, which makes it possible to assemble with hand tools alone.
A team of 4 can assemble a tank in a single day. The tanks are covered with a custom fabric roof mounted 
on a steel frame. 
The tanks are lined with a custom water-tight and light-tight bladder, which has mount points for the PMTs,
penetrations for cables and an access port for PMT installation and maintenance. Each tank 
contains 200,000 liters of purified water. The water is softened, filtered to remove particles
less than 0.5 microns and sterilized using UV light. The first tanks to be installed, which have been operational for 
3 years and we have noticed, show no indications of degrading water quality. Each tank contains 4 upward-facing PMTs mounted 
on the bottom. In the center is a high quantum efficiency 10''  Hamamatsu R7081-HQE PMT. It is surrounded by 3 8''
Hamamatsu R5912 PMTs, which were recovered from Milagro. 

Analog PMT pulses are connected to shaping electronics in the counting house through a 175m
cable path. The pulses are shaped and discriminated at 2 thresholds, at ~$\frac{1}{4}$ PE and ~5 PE, and threshold crossing
times are digitized using CAEN VX1190A multi-hit TDCs. All hits are recorded to memory and triggering 
is entirely done using software. The total data collection rate is ~400MB/s and drops to 15-20MB/s
with the application of a trigger. Events are reconstructed in real-time (few second latency) 
using an online analysis farm so that HAWC can rapidly alert other experiments of transient sources.

\section{Operation}

The deployment of the HAWC tanks took about 2.5 years to complete. During the deployment,
we were able to operate a subset of the full detector to verify the DAQ and to refine the 
deployment procedure. In September 2012, the first 30 tanks were completed and operated as an 
engineering prototype \cite{hawc30}. We incrementally expanded this array from 30 tanks to 
77 tanks to 95 tank and in August 2013, we began operation of a
111-tank array, herein referred to as HAWC-111. This array was operated until July 2014. 
The data presented in this proceedings by the HAWC collaboration
is mainly taken from the HAWC-111  operations period. Though HAWC-111 is 
considerably less sensitive than the full HAWC detector, it is about 3-5 times more sensitive
than Milagro and it is the most sensitive EAS gamma-ray detector ever operated. As a general
rule, the sensitivity of HAWC grows linearly with the number of tanks, so HAWC-111 is about 3 times
less sensitive than the full HAWC-300 detector.

On November 26, 2014, Thanksgiving day in the US, data taking began with 250 tanks. The US 
funding agencies only provided support for the completion of 250 tanks, though a 300-tank detector
was proposed and eventually completed with unallocated contingency. The detector was expanded
through the winter of 2014/2015 and full HAWC operations officially commenced at on March 20, 2015 at an
inauguration ceremony attended by officials of the US and Mexican funding agencies and the 
representatives from the scientific community.

The HAWC detector digitized and records all hits in all PMTs and a trigger was applied in the real time
processing system, permitting  the utilization of arbitrarily complex triggers.
HAWC-111 was operated with a EAS trigger rate of ~10 kHz. A trigger was initiated 
by the detection of coincident PMT hits in a 150 ns time window. Typically a threshold of
15-25 PMT was selected. Lower thresholds permit the detection of smaller events, lower
energy showers gamma-ray showers, but too low a threshold will introduce combinatorial  
backgrounds from uncorrelated hits. The near-threshold gamma-ray vents generally have poor angular 
resolution and are difficult to separate from background, so the sensitivity to hard
spectrum sources, such as the Crab, does not depend strongly on the choice of trigger threshold.

\begin{figure}
\centering
\includegraphics[width=2.7in]{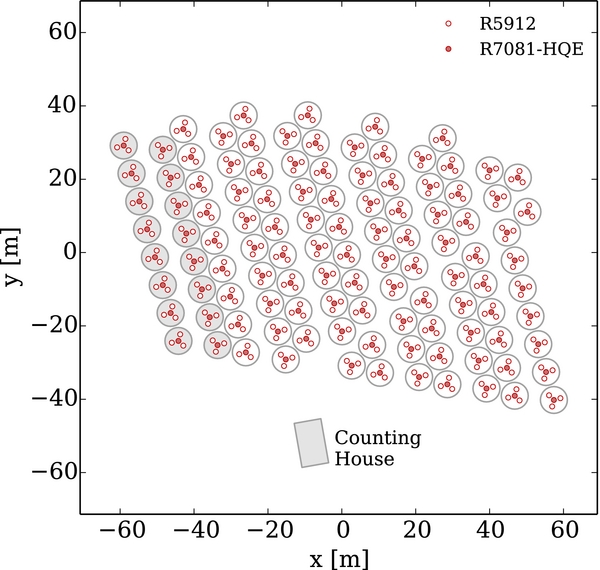}
\caption{Configuration of HAWC-111 array, which is roughly made up from the northeastern third of the full HAWC detector.}
\label{fig:hawc111}
\end{figure}

HAWC-111 was operated from August 2, 2013 to July 8 2014.  A total uptime of 83\%, which is a combination of detector off time
due to both operations interruptions and shutdowns to accommodate construction, and data lost due to analysis cuts. 

\section{Event Reconstruction}

The events are reconstructed by first fitting the spatial charge distribution to identify the 
location of the primary particle trajectory as projected onto the array, which we refer to as the shower "core", and then 
fitting the times of the hits to a shower plane 
hypothesis. An accurate core fit is important to properly account for the curvature of the
shower front and achieve the most accurate possible angle fit. The angular resolution of 
HAWC can be approximated as a 2D Gaussian with a width of $<0.2$ deg for events that hit 
nearly all the PMTs to $1-2$ deg for events near the trigger threshold. 

Generally, hadron-induced events have very different energy deposition patterns than those from
gamma rays. Specifically, hadrons are "patchy" with no distinct core while gamma-ray
showers are comparatively smooth with a pronounced core. Additionally, hadron induced events typically contain
muons, which can be identified by the large light depositions in the HAWC detector.
To reject events which don't have a gamma-like lateral distribution, we cut on the 
$\chi^2$ of the core fit. The core fit hypothesis is an NKG~\cite{nkg} function, which
is an analytical approximation of the lateral distribution of a gamma-ray shower. To identify muons,
we exclude events with large hits (typically $>~30$ PEs) found far ($>40m$) from the core. 
This cut reliably identifies muons. Though there is considerable overlap in the approachs
of the 2 cuts, they can be thought of as spatial cuts on the lateral distribution, the former
cut tests for consistency over the full shower profile and the latter searches for small 
spatial scale discrepancies. Application of these cuts can reject anywhere from 75\% 
of the background for near threshold events to $>$99.9\% for the largest events.
Significant improvement of the gamma/hadron separation capability is likely since the 
cuts shown here are rather rudimentary by design. Two articles in the proceedings 
address possible improvements in the pattern based separation~\cite{ICRCTomas}~\cite{ICRCZig}.

\section{Data Analysis}

Background estimation is performed using a technique called ``direct integration''~\cite{directintegration}. Since 
the field of view of the HAWC detector is very large compared to a gamma-ray 
source candidate, there is plenty of off-source data to be utilized for background 
estimation. Direct integration exploits the fact that the distribution of events on the 
sky in local coordinates ($\theta$,$\phi$) is highly stable, even when trigger rates change,
either do to operational changes or fluctuations in atmospheric pressure. Knowing this,
we can estimate the background for a specific astrophysical source location for a small time interval 
(small enough that the source can be assumed to be stationary) from the data collected 
prior to and after the source transited that location. These small intervals are integrated to
predict the background over the entire observation period.
Typically data from 2hr time blocks (30$^o$ of earth rotation)
are used to estimate the background for source regions of ~1$^o$, the background sample is 
30x larger than in the signal region. This technique has been shown to estimate backgrounds reliably to 
better than 1 part in 10$^4$.

\begin{figure}
\vspace{1in}
\centering
\includegraphics[width=5in]{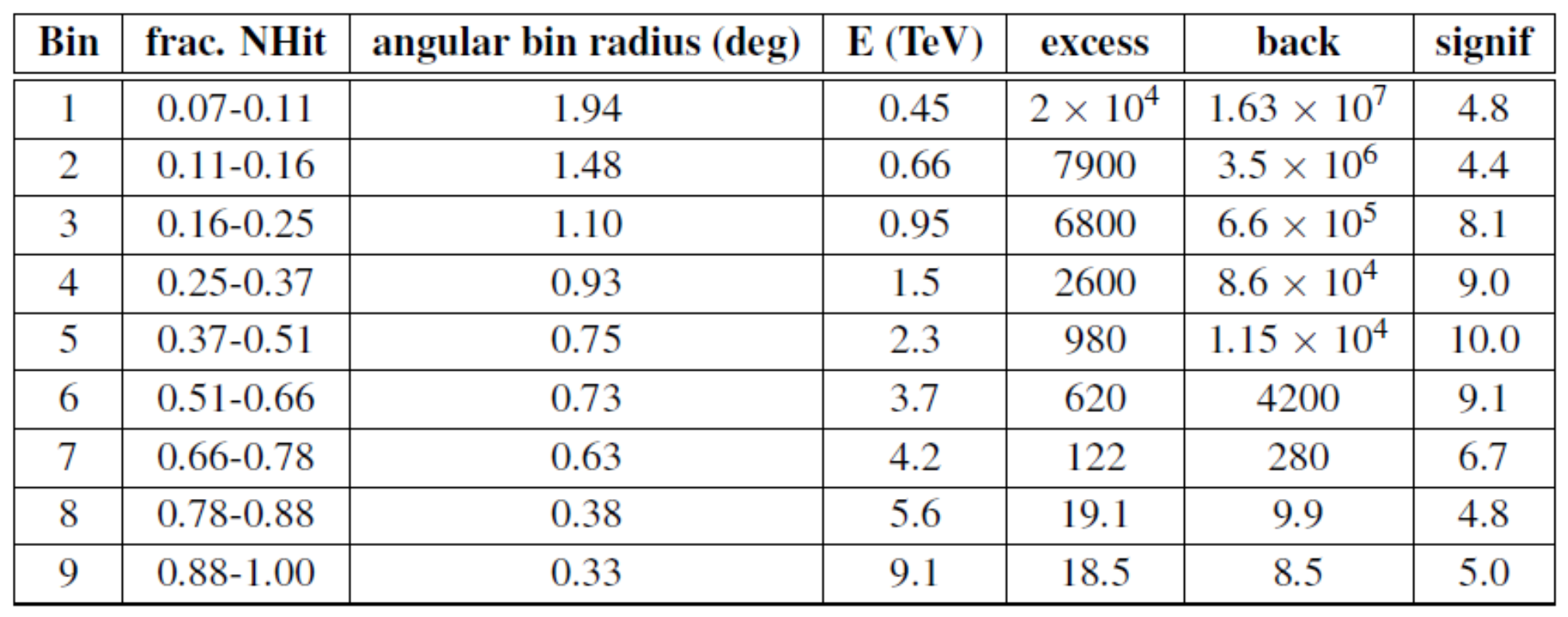}
\label{fig:hawc111_table}
\caption{Table showing the observed signal and background after cuts for the Crab for the nHit bins. Bin 0
was excluded since the trigger for HAWC-111 during portions of the operational period was too high to 
efficiently record events in this bin. Listed in the table are the bin number ({\bf bin}), the fraction of the 
PMTs hit for events in the bin ({\bf frac. NHit}), the median energy for the bin ({\bf E}), the detected gamma-ray 
excess and background ({\bf excess, back}) and the significance of the crab detection ({\bf signif}). Note that
within the HAWC-111 data set, multiple epochs were defined to account for the evolving detector. The data
shown in this table is for an epoch that includes only 180 days of the 283 days that were analyzed.}
\end{figure}

The analysis is optimized by separating the data into 10 independent ``nHit'' bins (labeled 0-9) ranging
from small events near threshold, which are abundant, in bin 0 to those with nearly every PMT recording
a hit in bin 9. The bins are defined so that the total hadronic rate is lower by a factor of 2x compared
to the preceding bin: e.g., bin 1 contains twice as many events as bin 2, which contains twice as 
many as bin 3 and so on. Bin 9 contains the largest 0.1\% of events by size.
While bin number correlates to energy, the bins are not a good energy estimator, since
shower energy depends also strongly on the core location and the zenith angle of the shower. However,
the \"nHit\" bin is a reliable predictor of the angular resolution and the optimal cuts for 
the suppression of the hadronic background. Figure \ref{fig:hawc111_table} shows a the measured
excess for each nHit bin for the HAWC-111 detector during it's 283 days of operation.
This methodology also used in the in the HAWC sensitivity paper \cite{sensi-paper}, only in the case of 
HAWC-111, we optimized the cuts on the gamma-ray signal from the Crab rather than simulation. For more 
this analysis, see \cite{ICRCPaco}.

\begin{figure}
\includegraphics[width=3in]{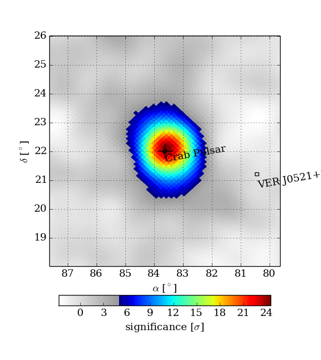}
\includegraphics[width=3in]{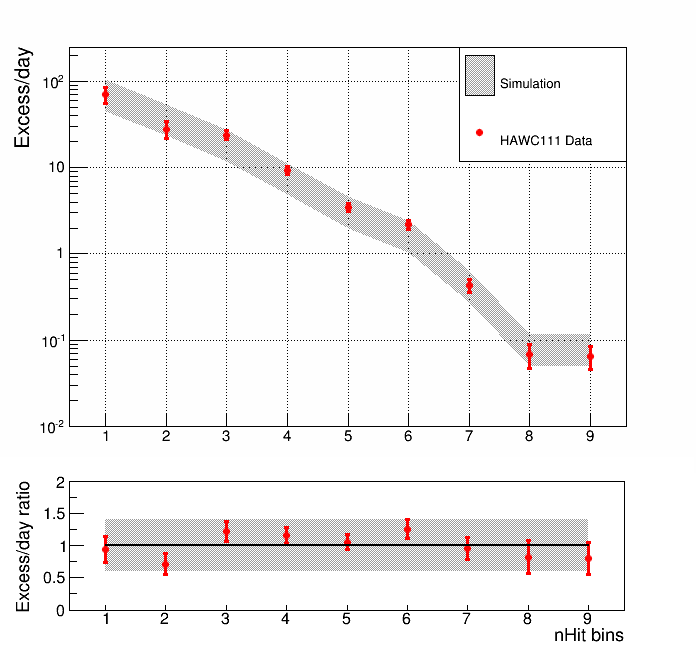}
\caption{Significance map showing the region of the Crab. HAWC-111 detected the Crab at 24.5$\sigma$, ~1.5$\frac{\sigma}{\sqrt{day}}$ (left).
Figure showing the excess measured for each of the nHit bins and the predicted value from a full detector simulation (grey band). A 40\%
              systematic error is assumed, which dominates the error in the predicted event rates (right).}
\label{fig:crabbins}
\end{figure}

We combine the bins to form a single significance estimate for the excess using 2 methods: 1) a Poisson
Likelihood based on the expected rate for a hypothesis spectrum and 2) by weighting the events from each bin
with a value proportional to the signal to background ratio. When the number of events is large, the 2 methods
give identical results. Figure \ref{fig:crabbins} shows the significance map in the 
vicinity of the Crab and the bin-by-bin excess compared to simulation assuming the Crab flux and spectrum as
measured by HESS~\cite{hesscrab}. 

\section{Importance of Gamma-Hadron Separation}

Unlike many previous EAS detectors, HAWC has the ability to efficiently separate gamma-ray showers from 
the isotropic hadronic background. The utilization of the water Cherenkov technique for mapping out the 
lateral distribution of shower energy and identifying muons was pioneered by the Milagro experiment, 
but drastically improved by HAWC. This improvement mainly comes from the much larger area covered by 
deep water, which is nearly 10x greater in HAWC. 
Prior to the completion of the HAWC detector, we published a paper on the gamma-ray sensitivity of 
HAWC. At high energies, >~10 TeV, the HAWC detector was so efficient at rejecting hadrons 
that it was computationally difficult to simulate enough events to measure the passing fraction. 
Consequently, we only published limits on our sensitivity in this portion of the energy band.
As HAWC has been completed, it is now possible to directly measure the background rates using
data and we have found that for events which hit >75\% of the PMTs, hadrons can be rejected with rates
as high as 99.99\% while retaining more than 30\% of gamma-ray showers. These event sizes roughly 
correspond to the energy band $>$10 TeV. Data demonstrating the background rejection of the full
HAWC detector will be presented at the meeting.

This capability will give HAWC unrivaled sensitivity for deep sky surveys at the high end of the TeV 
range. Though HAWC's effective area is only about ~20,000 m$^2$, in 5-years of observing, it will 
have an integral exposure of 10$^8$ m$^2$ hrs for sources in the observable sky (-25<$\delta$<65).
This exposure will surpass all but the deepest observations of CTA.

\section*{Acknowledgments}

\footnotesize{
We acknowledge the support from: the US National Science Foundation (NSF);
the US Department of Energy Office of High-Energy Physics;
the Laboratory Directed Research and Development (LDRD) program of
Los Alamos National Laboratory; Consejo Nacional de Ciencia y Tecnolog\'{\i}a (CONACyT),
Mexico (grants 260378, 55155, 105666, 122331, 132197, 167281, 167733);
Red de F\'{\i}sica de Altas Energ\'{\i}as, Mexico;
DGAPA-UNAM (grants IG100414-3, IN108713,  IN121309, IN115409, IN111315);
VIEP-BUAP (grant 161-EXC-2011);
the University of Wisconsin Alumni Research Foundation;
the Institute of Geophysics, Planetary Physics, and Signatures at Los Alamos National Laboratory;
the Luc Binette Foundation UNAM Postdoctoral Fellowship program.
}

\end{document}